# A Smart City Model for an Intelligent Traffic Light Decision System

Darko Pajkovski, MarijaApostoloska – Kondoska and HristinaDimova Popovska

Faculty of Information and Communication Technologies, University "St. KlimentOhridski", Bitola, Republic of North Macedonia

*ABSTRACT*

*A smart city is a framework that uses information and communication technologies to improve public safety, quality of life, transportation and energy efficiency. A big share of these technologies has intelligent networks consisting of connected objects and devices that transmit data using wireless technology and cloud-based solutions. These technologies and their applications can receive, analyze and manage data in real-time to help citizens and municipalities make better decisions to improve quality of life. Also, pairing devices and data with already existing physical infrastructure can cut costs and significantly improve sustainability. The main focus of this work is managing traffic flows in urban transport networks regarding minimizing traffic congestion and improving air quality. The main goal of the proposed models is to provide a good starting point for the future development of smart city models and their implementation in urban areas that show significant congestion and air quality problems.*

*KEYWORDS*

*Smart city, air pollution, traffic congestion, IoT, PLC, decision support systems.*

## 1. INTRODUCTION

The total number of vehicles worldwide has grown rapidly in the last few years. Road safety has become one of the main issues for governments and car manufacturers in the past decade. With the development of new technologies, researchers and institutions had to focus their efforts on continuously improving road safety. The evolution of Wireless Sensor Networks (WSN), has allowed researchers to design and develop communications systems and models where vehicles directly take part in the network [8].

Nowadays, half of the world's population lives in urban areas [21]. Most of them travel, thus they are facing many challenges related to transport. From the point of view of transport management, the results of road congestion increase operating costs and decrease the efficiency of transport management. The effects caused by transport congestion also affect social factors, air quality, accidents, noise pollution, and other issues that degrade the whole environment. Transport-related emissions are the biggest source of air and noise pollution. The primary air pollutants emitted from motor vehicles are carbon monoxide (CO), particulate matter (PM), and nitrogen oxides (NO). The issue of health risks posed by the transport sector become a global environmental concern. According to the World Health Organization, Republic of North Macedonia [13] is one of the countries most affected by air pollution in the world and the rate of premature death is much higher than in most EU countries. Breathe-Life reports that every year 2.574 people die prematurely as a direct result of air pollution. Skopje, Tetovo and Bitola are the cities with 4-5 times higher concentrations of particle PM 2.5. Moreover, the winter temperatures are making the

 

situation worse, because of the emissions resulting from the use of fossil fuels for heating residential and commercial structures.

Therefore, the average low income and energy poverty are two main issues making the capital city in the Republic of Macedonia - Skopje such a polluted city. Also, Tetovo and Bitola are dynamic cities with longer travel times, queues, high energy consumption and air pollution. In combination with all the above and the dense concentration of administrative and other service activities in the city's main core traffic congestion and air pollution are just two of the issues making these cities such polluted. Another reason is the natural position of the capital city - Skopje located in a valley surrounded by mountains that trap the fog. Temperature invasion additionally complicates the case like a natural phenomenon which causes warm air to remain above cool air and contributes to the greenhouse effect [1][2] .

In 2015 innovative web and mobile application called "MojVozduh" or "AirCare" was developed. Application data analysis shows that air pollution in Macedonia is 20 times over the EU threshold and four times more than in Beijing [13] .

This paper focuses on creating an adaptive signal control system for traffic lights at intersections experiencing severe congestion. It outlines two key methods for alleviating traffic jams and reducing air pollution in suburban or rural areas. Using image processing techniques combined with real-time traffic data at the intersection, we propose a smart traffic light system that can adjust the timing of light changes based on the volume of vehicles on the road.

The rest of the paper is structured as follows. The second section describes related works, the third section identifies road traffic, the fourth section is our model of an intelligent traffic decision system developed for reducing traffic congestion, and the last section concludes the previously described work.

## 2. RELATED WORK

Recently, traffic congestion has become a major and common problem in more countries, especially in cities with an increased population density the number of cars is automatically high. The authors in [10] have presented several simulations based on symmetry models implemented in such a practical case to streamline vehicle density and reduce traffic congestion. The authors in [11] have presented a comprehensive analysis of the most recent deep reinforcement learning (DRL) approach used for the developed algorithm which is applied to large traffic networks. Therefore, they provided best practices for choosing adequate DRL, parameter tuning, and model architecture design. Another approach connected to reducing traffic congestion and modeling smart traffic light systems is presented [12]. The authors have developed a simulator in which a grid of traffic lights is constructed at intersections and the flow of vehicles through the grid is optimized by developing algorithms for traffic light state change. A similar solution is presented by the authors of [3] demonstrated machine learning algorithms to predict traffic conditions and traffic light timings. Simulation results indicate the dynamic traffic interval technique is more efficient with a reduction in waiting time of vehicles from 12% to 27% compared to the fixed time traffic light control methods.

This study [14] introduces a groundbreaking integrated approach designed for smart cities, focusing on fostering environmentally sustainable economies through innovative technological and socio-economic transitions. The model developed in this research calculates the smart city index (SCI) by consolidating 32 unique performance indicators. These indicators collectively drive significant improvements across various sectors, including the environment, economy, energy, social aspects, governance, and transportation. The proposed SCI model provides a comprehensive framework for evaluating and enhancing the multifaceted dimensions of smart city development, ensuring a holistic advancement towards sustainability and smart urban growth. The model is applied to 20 selected cities worldwide, identifying Sydney, Osaka, Toronto, Montreal, and Vancouver as leaders in smart and innovative urbanism, achieving the highest scores among the 20 cities analyzed. These cities excel in integrating technological advancements and socio-economic transitions, positioning themselves at the forefront of sustainable and smart city development.

The authors in [15] have focused on extracting insights from city data, beginning with the research design and progressing to recommendations for data-driven smart city solutions. Consequently, the study also investigates how various machine learning and deep learning techniques can be utilized to develop analytical models in data-driven smart cities, aiming to meet the needs of the populace. This article [16] argues that city councils can employ a business model logic to design and implement smart city services. It demonstrates specifically how a framework for designing city business models can be applied during the planning phase of an integrated ICT city platform. The authors of this paper [17] propose a new system that integrates the Internet of Things (IoT) with a predictive model based on ensemble methods to optimize the prediction of parking space availability in smart parking systems. Authors of the [18] first presented a study of smart definitions and domains. Therefore, they review and highlight the technical standards for smart cities implementation. Moreover, they observed that healthcare (23% impact), mobility (19% impact), privacy and security (11% impact), and energy sectors (10% impact) have a more significant influence on AI adoption in smart cities. This review paper [19] aims to benefit government officials, businesses, and researchers who want to extend smart city development. Therefore, they discuss the challenges the developers generally face while implementing smart city applications. The goals of this systematic review [20] were to explode the involvement of surveillance drones in smart cities. Findings show that surveillance drones were used in seven distinct research fields (transportation, environment, infrastructure, object detection, disaster management, data collection and other applications). Therefore, air pollution and traffic monitoring were the dominant application areas.

Authors of [22] proposed a solution with an integrated camera IoT sensor in every traffic signal corner to monitor the vehicle flow, processing and optimizing vehicle flow before sending it to the cloud processes. Data from the various signal corners runs an algorithm to detect traffic direction and controls the signal lights.

## 3. IDENTIFICATION OF ROAD TRAFFIC

With the rapid development of technology, the total number of vehicles significantly increases. The result is traffic congestion and frequent accidents in major cities. Approximately one-third of the total population of the Republic of North Macedonia is concentrated in Skopje. The existing traffic network is a combination of ring roads, intersections with traffic lights and other radial and orthogonal infrastructure segments [6][7].

This paper aims to develop an adaptive signal control strategy for traffic light intersections with heavy traffic congestion. Sometimes, vehicles traversing the primary routes endure protracted waits. It is very often to wait for several cycles of traffic lights to pass the crossroads meanwhile on the opposing side vehicle presence remained sparse. This approach is a great waste of the owner's time and road resources as well as the non-utilization of full road capacity. Based on the above reasons, we will describe two fundamental approaches to reducing traffic congestion and air pollution in extra-urban areas. With the image processing technique and traffic flow at the intersection, we proposed an intelligent traffic light model able to delay its state depending on road vehicle flow.

## 4. MODEL OF AN INTELLIGENT TRAFFIC DECISION SYSTEM

Nowadays, most of the traffic lights in the Republic of North Macedonia are controlled by a single chip with fixed periods for red, yellow and green lights respectively. Of course, this kind of system has been made by multiple researchers in past about main street congestion, but over the years the infrastructure of the city has changed resulting in the required update of the traffic lights system. Therefore, this paper proposed PLC (Programmable logic controller) usage to control the traffic light signal. According to the authors' driving experience, the conversion time of traffic lights in Skopje is between 10 and 90 seconds. The specific classification of green light delay period and vehicle traffic situation is shown in Table 1. The main points of developing this intelligent traffic light decision system are divided into three grades 1, 2 and 3. All of them are used to create a condition of the branch statement. With the results of this statement, all traffic lights are delayed respectively. Red light is delayed by 10 seconds when the traffic is low, 30 seconds when the traffic is large and 60 seconds when the traffic is huge. Table 1 Traffic light signal

|   | Green light delay period | Vehicle traffic | Traffic congestion |
|---|---|---|---|
| 1 | 10 to 20 seconds | 5 to 8 cars | Low traffic flow |
| 2 | 20 to 40 seconds | 8 to 15 cars | Large traffic flow |
| 3 | 40-80 seconds | 15 to 30 cars | Huge traffic flow |

There are a few approaches for measuring the road traffic of vehicles divided into two categories [4]. The first one is sensor detection a completely embedded system based on an electric circuit. Inductive sensors are designed to detect the moving objects. Simply, the inductive sensor is a wire which is placed under the moving objects and directly connected to the controller (PLC). One of the main disadvantages nowadays is that this method requires the destruction of the road surface. Implementation of this type of solution is significantly complicated and it is very hard to maintain nowadays. Also, the other type is the suspension system including infrared and radar detection systems with high cost of implementation and single collection of information. Laser radars or Lidars use the technology of gathering and processing data from remote objects using optical systems using light reflection and dispersion in transparent media.

The accuracy of this type of laser seriously depends on weather conditions (rains and snow can influence its outfeed).

The global positioning system (GPS) can be used for traffic flow data collection. Every vehicle that uses GPS or in which there is a mobile phone or other GPS device acts as a sensor moving along the complete road network. All data collected (location, speed and moving direction) can be sent to the central processing center. This database can be used for developing new methods for calculating the speed, travel time and therefore the performance of the complete system from the vehicle's data on the road. This approach needs more time for data analysis and computing therefore data accuracy can vary on the quality of the data sources (different types of GPS systems).

With the rapid development of computer vision algorithms and image processing technology, vehicle flow detection, highway lane following systems and lane-keeping assist become the trend of vehicle systems that allow transportation and energy efficiency [9]. In this work, we are focused on vehicle flow detection technology and its advantages of large coverage, a huge amount of detection information and convenient maintenance. Also, this kind of technology is simple to install, because of newly developed traffic lights, and camera capture lenses are mostly embedded. Nowadays, creating intelligent traffic lights with computer vision and image processing technology only needs to solve the software issues because the hardware no longer needs additional supplies resulting in great savings. A roadway intersection with traffic lights in Skopje is shown in Fig. 1.

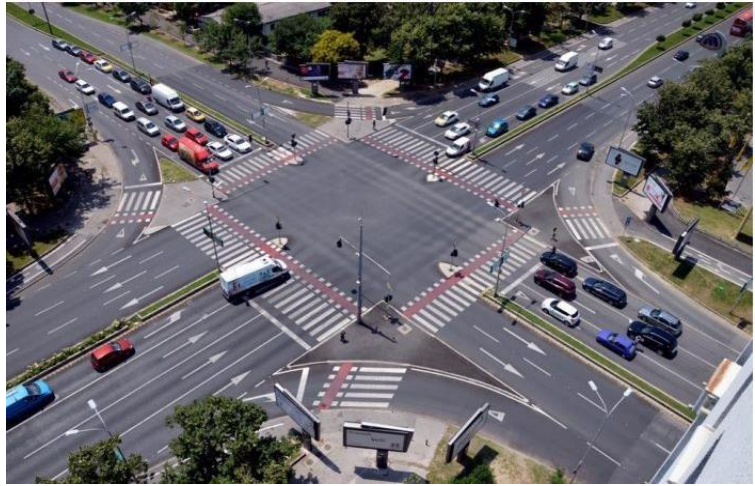

Figure 1 – Roadway intersection with traffic lights in Skopje

One picture directly from the camera cannot be used to read traffic flow. The reasons are:

- The image definition from the camera sensor is not high.
- Different times of day and different periods of day result in different light intensity and data processing output would be non-consistent.
- Different colors on the road (dark and light) may influence the optical potential. Lighter color means stronger photo potential.
- Capture picture angle may result in negative effect.

Because of all the reasons described above, we may conclude that it is very difficult to read useful information directly from a camera. As we know for the influence of temperature on triode [5] the common emitter amplification circuit composed of triode very often causes signal distortion due to the instability of the static working point. So, we are using a differential circuit to eliminate the zero drift of the whole electronic circuit. Now we can compare the values of the two roads. As a result, we get the relative congestion of the road and then we are computing traffic light delay. From Figure 1, the direction east-west is X and north-south is Y. The whole system of data processing with computer vision technique may be organized as follows: 1) Read the camera sensor and take photos 2) Apply grey level and threshold processing 3) Extract X and Y separately with vehicle flow density (respectively Kx and Ky) and compare. Comparing X and Y direction is given as follows:

- $\frac{Kx}{Ky} \geq 3$; red light should be delayed for 60 seconds.
- $2 \geq \frac{Kx}{Ky} < 3$; red light should be delayed for 30 seconds
- $1 \leq \frac{Kx}{Ky} < 2$; red light should be delayed 10 seconds.

Through the above analysis we wrote the flow chart of the program to control the intelligent traffic decision model shown in Figure 2. The realization of the intelligent traffic lights system utilizes real-time data and advanced algorithms to optimize traffic signal timings, and dynamically manage traffic flow, reducing congestion and delays on the same time enhancing overall road safety

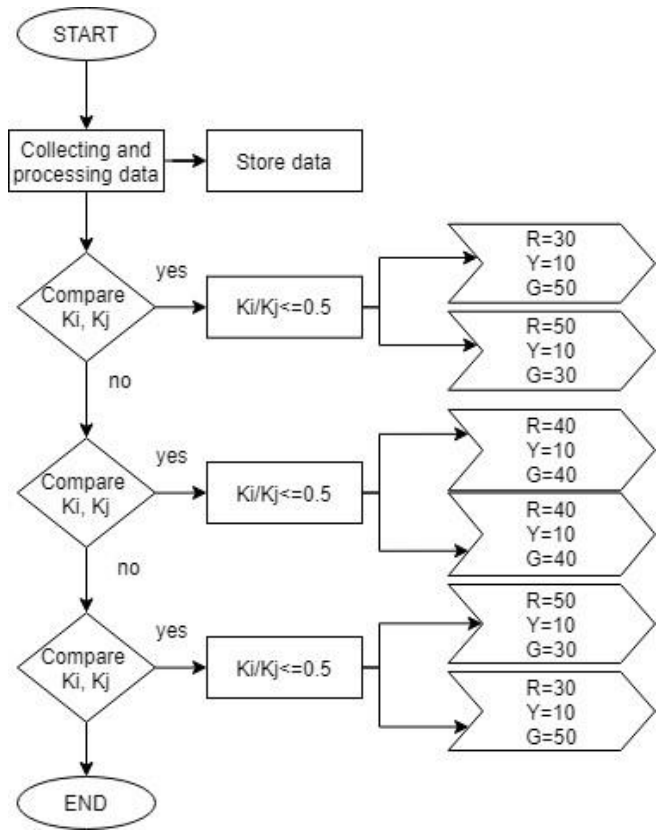

Figure 2 Flowchart of intelligent traffic decision model

## 5. CONCLUSION

In this paper, the traffic congestion caused by various reasons was analyzed and presented traffic light decision model. These elements were made based on a real case and based on the information received on real conditions in our country. Reducing traffic congestion is especially important for urban areas where the air quality is mostly degraded due to various types of pollutants emitted from motor vehicles.

The realization of an intelligent traffic decision model can make the traffic flow in all directions achieve a reasonable state. There are direct and indirect benefits of reducing traffic congestion like increasing the number of parking, improve traffic safety and energy efficiency. Therefore, this system enables more efficient use of transportation infrastructure and resources by optimizing traffic flow, reducing road idle time and minimizing fuel consumption, resulting in cost savings for all road users. Moreover, it can effectively protect the natural environment that people depend on and create a beautiful trip environment for people.

Overall, the proposed solution of an intelligent traffic decision system offers numerous benefits, ranging from improved traffic management and safety to environment sustainability and economic development, contributing to the creation of smarter, more efficient and more livable cities. In other words, traffic light decision systems play a crucial role in managing traffic flow, enhancing safety, reducing environmental impact, and improving overall transportation efficiency in urban areas.

## AUTHORS

**Darko Pajkovski** is a Master of Computer Science and Engineering at Faculty of Information and Communication Technologies, University St." KlimentOhridski", Bitola, Republic of North Macedonia. He is interested in digital design, IoT, computer graphics, AI etc.

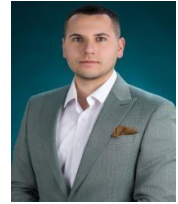

**Marija Apostoloska – Kondoska** is working as a Assistant at Faculty of Information and Communication Technologies, University St. "KlimentOhridski", Bitola, Republic of North Macedonia. Her qualifications are M.E. Softwere application engineering, Ph.D. (pursuing) and having more than 10 years of experience in public institutions and private company and 2 years of teaching experience.

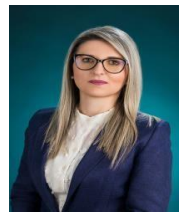

**Hristina Dimova Popovska** is a Master of Information Sciences at Faculty of Information Sciences in Bitola, Macedonia. She is interested in computer graphics, internet of things etc.

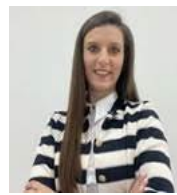